# Generation of micro-Joule level coherent quasi-continuum extreme ultraviolet radiation using multi-cycle intense laser-atom interactions

Vassilis Tsafas [1†], Theocharis Lamprou[1,2†], Emmanouil Skantzakis [1*], Arjun Nayak[3,4], Dimitris Charalambidis [1,2,3], Paraskevas Tzallas [1,3] and Ioannis Orfanos[1*]

[1]*Foundation for Research and Technology - Hellas, Institute of Electronic Structure &Laser, PO Box 1527, GR71110 Heraklion (Crete), Greece.*

[2]*Department of Physics, University of Crete, PO Box 2208, GR71003 Heraklion (Crete), Greece*

[3]*ELI-ALPS, ELI-Hu Non-Profit Ltd., H-6720 Szeged, Hungary.*

[4]*University of Szeged, Dugonics tér 13, H-6720 Szeged, Hungary.*

[†] Equally contributing authors.

*Corresponding authors e-mail addresses: orfanos@iesl.forth.gr , skanman@iesl.forth.gr

## Abstract

In the present work we report on the current progress of the recently constructed GW attosecond extreme ultraviolet (XUV) source developed at the Institute of Electronic Structure and Laser of the Foundation for Research and Technology-Hellas (I.E.S.L-FO.R.T.H.). By the implementation of a compact-collinear polarization gating arrangement, the generation of a broadband, coherent XUV quasi-continuum produced by the interaction of a many-cycle infrared field with a gas phase medium is achieved. The spectral width of the XUV emission generated in Xenon, is spanning in the range of 17-32 eV and can support isolated pulses of duration in the range from 0.4 *fs* to 1.3 *fs* and pulse energy in the 1 μJ level. Theoretical calculations, taking into account the experimental conditions of this work, are supporting the observations, offering also an insight regarding the temporal profile of the emitted radiation. Finally, the high intensity of the produced XUV pulses has been confirmed by investigating the two-XUV-photon double ionization process of Argon atoms. The demonstrated results inaugurate the capability of the beamline to produce intense isolated attosecond pulses towards their exploitation in studies of non-linear XUV processes, attosecond pulse metrology and XUV-pump–XUV-probe experiments.

# 1. Introduction

The observation of ultrafast motion and dynamics in the microscopic scale in all states of matter is fundamental towards our understanding of the processes governing the nature. Tools for tracking down such dynamics, have proven to be bursts of light with durations less than the characteristic time scales of the phenomena under investigation. Since the early 90's laser driven gas phase high harmonic generation (HHG) **[1,2]** attracted tremendous interest due to the capability of producing light bursts with attosecond (1 attosecond (*as*) = $10^{-18}$ sec) pulse duration when the generated harmonics are phase correlated, as theoretically predicted **[3,4]**. In fact, during the last two decades, progress in pulse engineering and the rapid development of high-power laser systems allowed the generation of attosecond pulse trains (APTs) and isolated attosecond pulses (IAPs) which are now routinely produced from gas phase targets **[5]** (and references therein). APTs and IAPs are created by the superposition of a high harmonic (HH) frequency comb and continuum (or quasi-continuum) frequency modes respectively, laying in the extreme ultraviolet spectral range (XUV). Such pulses pose as unique contrivance for real time observation of electronic motion **[6-11]** (and references therein), especially when sufficiently intense to induce non-linear process in the XUV region **[12-28],** offering thus the opportunity for their exploitation in XUV-pump-XUV-probe schemes **[29-32].** The key feature in all cases that succeeded in generating intense XUV pulses, is geometries where the driving laser field is loosely focused in the HHG medium **[33-35]**. Such configurations have been implemented in the past for the generation of high photon flux XUV pulses **[16-27]**. Concerning intense APTs, the latest significant achievement is the generation of XUV APTs of energy up to ~200µJ per pulse **[22,23]**. The requirement of the generation of intense XUV pulses in the form of few-pulse APTs or IAPs can be achieved by means of high power few-cycle driving laser fields, with the latest findings to be the generation, in Argon gas, of approximately 1 µJ sub-*fs* pulses. **[21,24-27]**. An alternative strategy is the exploitation of high-power multi-cycle driving laser fields combined with techniques known as Polarization Gating (PG) **[36-39,42]** or Ionization Gating (IG) **[40,41]**. The former approach leads to the development of XUV beamlines that have now succeeded in generating intense IAPs with ≤1 *fs* duration in the nJ range **[42-44]**.

In the present work, we demonstrate the generation of ~ 1.3 µJ quasi-continuum XUV spectra that can support sub-fs XUV pulses of carrier wavelength approximately 61 nm. This has been achieved by developing a compact-collinear, multi-cycle polarization gating (CCMC-PG) hosted in the recently developed GW-scale XUV beam line of FORTH **[22,23]** using Xenon gas as HHG medium. Theoretical calculations, which have been performed taking into account the experimental conditions, support the experimental results and allow us to estimate the temporal confinement of the XUV pulses, which was found to be in the range of 0.4 *fs* to 1.3 *fs*. Such XUV pulses in tight focusing geometries can support intensities in the range of $10^{14}$ W/cm$^2$, sufficient to induce non-linear processes. This has been confirmed by measuring the two-XUV-photon double ionization process of Argon atoms.

## 2. Experimental set-up and methods

The experimental set-up implemented in this work is illustrated in Fig. 1(a). The studies were performed at the recently constructed 10 Hz repetition rate GW-scale XUV source of the Attosecond Science and Technology laboratory at FO.R.T.H.-I.E.S.L. The source is based on the gas-phase HHG process driven by high power p-polarized 24 fs, 800nm pulses, which are loosely focused, using a 9m focal length spherical mirror, into a Xenon gas. In previous works it is shown that the source can deliver intense APT with a ~10 *fs* long envelope and average pulse duration of 650 *as*, with energy >100µJ per pulse in the 17-30 eV photon energy range [23]. In the present work the p-polarized IR beam of ~70 mJ and 3 cm diameter was passing through the CCMC-PG optics. This creates an ellipticity-modulated driving field in time, in a similar way as is described in the work of ref. [45]. The CCMC-PG optics are placed before the 9 m focal length spherical mirror, while a super-Gaussian beam stop creates an annular shape in the IR beam. The CCMC-PG arrangement, is shown in Fig.1 (a) and consists of a zero-order half wave plate (ZO-λ/2), a multiple order quarter wave plate (MO-λ/4), a zero-order quarter wave plate (ZO-λ/4) and a polarizing beam splitter (PBS). The ZO-λ/2 waveplate with its optical axis (OA) at ~ 22.5° is used to rotate the polarization of the input pulse to 45°. Then the laser pulses pass through a MO-λ/4 waveplate with its OA at ~ 0°. This birefringent plate creates two delayed and partially overlapping pulses. The delay between the pulses is defined by $δ = d (1/v_e − 1/v_o)$, where *d* is denoting the thickness of the plate and $v_o$, $v_e$ are the velocities of the light along the ordinary and extra-ordinary axis of the plate, respectively. The delay between the two pulses introduced by the MO-λ/4 waveplate was carefully chosen to be $δ ≈ 24$ *fs* matching the laser pulse duration. After passing through the MO-λ/4 waveplate, the elliptically modulated IR field has circular polarization at the center of the overall pulse and linearly polarized at the edges. Then a ZO-λ/4 with its OA at ~ 135° waveplate converts the circularly polarized field, at the center of the pulse, to linear and the linearly polarized field at the tails to circular. Finally, a PBS with variable *s/p* transmission ratio regarding the field polarizations with respect to incident angle, is introduced in order to provide an accurate control over the amplitudes of the two perpendicularly polarized fields synthesizing the ellipticity-modulated IR pulse. In this way, a narrow temporal gate with linear polarization is formed within the pulse around its center. The gate width, $τ_g$, is given by the expression $τ_g ≈ [log_2(A) τ_L^2] / 2 δ$, where $τ_L$ is the initial pulse duration, *δ* the delay between the pulses and *A* a parameter linked with the amplitudes of the perpendicularly polarized $E_p$ and $E_s$ fields reading: $A = [−2R\sqrt{1 − B^2} + B(R^2 − 1)]/(B − 2R + BR^2)$ [43,45]. *R* denotes the amplitudes' ratio, while $B = \sin [2 \tan^{-1}(ε_{th})]$, where $ε_{th}$ being the ellipticity threshold which corresponds to the ellipticity value at which the harmonic generation yield drops to 50% of its maximum value.

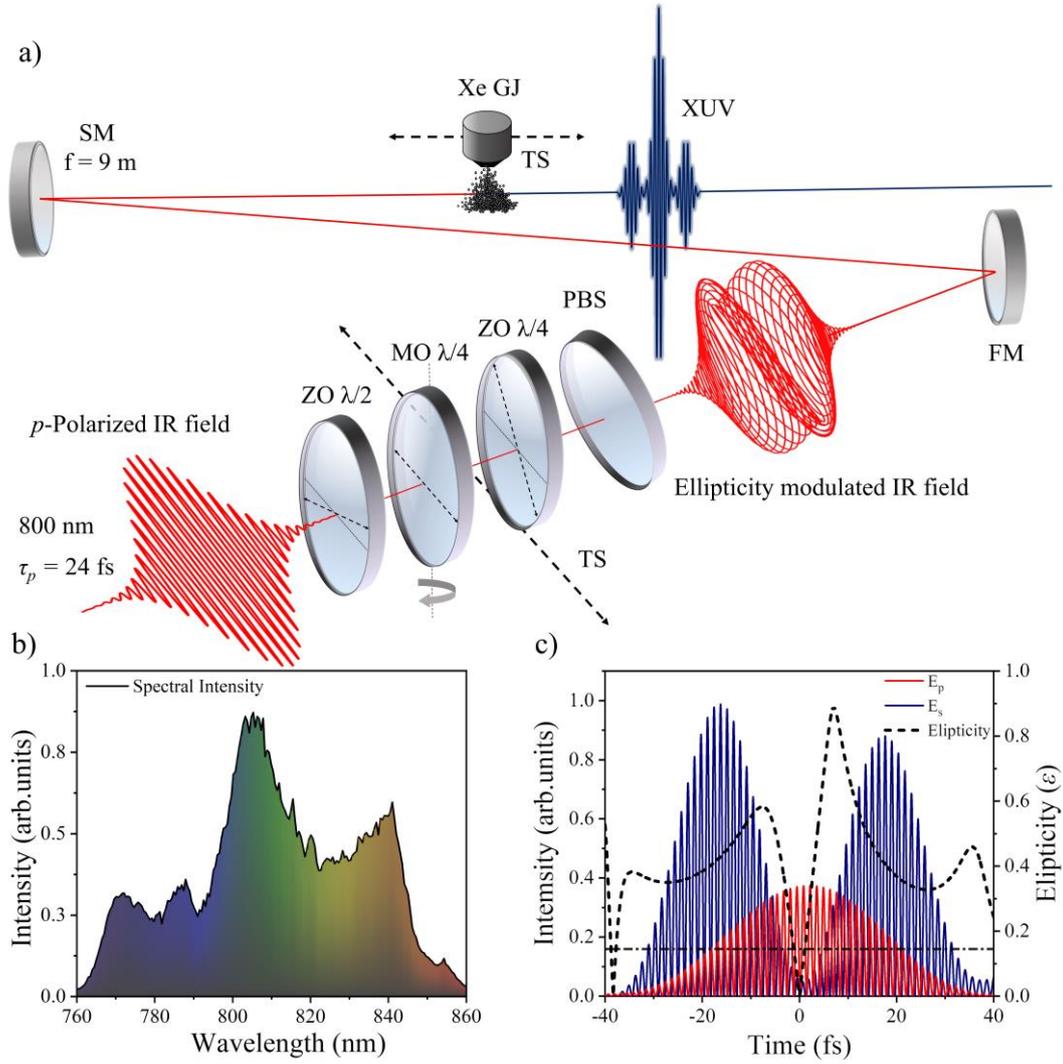

**Fig. 1. The compact-collinear, multi-cycle-polarization gating apparatus:** (a) a drawing of the CCMC-PG configuration. ZO λ/2: Zero order half wave plate, MO λ/4: Multiple order quarter wave plate, ZO λ/4: Zero order quarter wave plate, PBS: polarizing beam splitter, FM: flat mirror, SM: spherical mirror (The FM and SM mirrors have been placed very close to the normal incidence with respect to the incoming beam). Xe GJ: pulsed-jet filed with Xenon, TS: translation stage, used to move in- and out the CCMC-PG system from the beam path and the pulsed-jet filed with Xenon along the propagation axis, (b) Measured spectral intensity of the IR driving field. (c) dashed line: calculated ellipticity (ε) of the polarization-modulated pulse used for the generation of the quasi-continuum XUV radiation. Solid red and blue lines: calculated intensity of the two fields $E_p$ and $E_s$ respectively, of the ellipticity-modulated pulse. The ellipticity threshold $\varepsilon_{th}=15\%$ (dashed-dot black line) and the time gate width $\tau_g \sim 2\,fs$ are also shown.

In order to evaluate $\tau_g$, and the shape of the ellipticity-modulated driving field in the interaction region, calculations have been performed regarding the polarization gating conditions. Thus, the pulse duration of the driving field, the spectral phase distribution within the bandwidth of the driving field, the dispersion of the optical elements used, as well as the phase retardation of each frequency component introduced by the fast and slow OA of the MO-λ/4 waveplate during propagation have been taken into account. Fig. 1 (c) presents the calculated ellipticity-modulated IR field. The solid red and blue lines are representing the intensity of the two perpendicularly polarized fields $E_p$ and $E_s$ respectively, constructing the ellipticity-modulated driving pulse. The black dashed line shows the ellipticity ($\varepsilon$) of the polarization-modulated pulse used for the generation of the quasi-continuum XUV radiation. The threshold ellipticity $\varepsilon_{th}$=15% forming a time gate width $\tau_g$~2 *fs* is presented with the horizontal black dashed-dot line.

After the CCMC-PG arrangement, the IR pulses are focused by means of a 9m focal length spherical mirror into a Xenon gas jet. The generated XUV radiation and IR co-propagate and impinge on a Silicon plate (Si) placed at Brewster angle (~75$^0$) for the p-polarized part of the IR field. It is noted that the Si plate significantly reduces the amount of the p-polarized component of IR field and reflects 50-60% of the XUV beam energy [46]. An aperture of 5 mm diameter is placed after the Si reflection, blocking the annular IR beam created by the super-Gaussian beam stop. Furthermore, a mount hosting metallic filters is introduced in the beam path for the spectral selection of the XUV radiation, as well as to eliminate any residual part of the IR beam reflected by the Si plate. In this experiment, a 150 nm thick Aluminum (Al) filter (with approximately measured transmittance of 15 %) was used accompanied with an additional pinhole of 3 mm. A calibrated XUV photodiode ($PD_{XUV}$) (OptoDiode AXUC100G) mounted on a translation stage placed after the metal foils, was introduced in the beam path for measuring the XUV pulse energy. At the same chamber the beam profile was recorded by introducing an XUV beam profiler ($BP_{XUV}$), consisting of a pair of multichannel plates (MCPs) and a phosphor screen followed by a CCD camera. The transmitted XUV beam then reaches the end station in which the spectral characterization of the harmonic radiation takes place. The end station consists of an Argon gas jet and a magnetic bottle time of flight (MB-TOF) spectrometer, which operates either for energy resolved photoelectron (PE) mode or ion mass spectrometer. A gold (Au) spherical mirror, of approximately 12% reflectivity, mounted on a multiple-translation-rotation stage was introduced in the beam path in order to focus the XUV radiation into the Argon gas jet.

The spectrum of the generated XUV radiation was obtained by measuring the single-photon ionization PE spectra induced by the XUV radiation. For these measurements, the XUV spherical mirror was moved out from the beam path. The ionization products are originating solely from the interaction of the incoming-unfocused XUV beam with the Argon gas. Fig. 2 shows the recorded XUV spectra. The black-grey filled curve presents a typical HHG spectrum produced by the non-linear interaction of linearly *p*-polarized driving filed with Xenon. The blue curve is depicting the XUV quasi-continuum spectrum when the CCMC-PG system was introduced. The harmonic generation process then is restricted to ~ 1 laser cycle leading to a quasi-continuum XUV spectrum spanning in the range of 17-32 eV supporting IAPs with minimum duration of ~370 *as*. However, it should be noted that the generated XUV spectrum fluctuates, between continuum and discrete harmonics depending on the shot-to-shot CEP variation of the non-CEP-stabilized driving laser field. Thus, only a fraction of the laser's shots generates pure XUV continuum spectra corresponding to

IAPs. In fact, due the strong dependence of the generated XUV spectrum on the CEP values, the equivalent product in the time domain varies between single IAP and narrow few-pulse APT, a topic which will be addressed in the following section. The insets (a) and (b) of Fig.2 are presenting characteristic $PD_{XUV}$ signals, corresponding to the XUV pulse energy in the case of producing a HH frequency comb and quasi-continuum spectrum respectively. For the energy measurements, in both cases, the $PD_{XUV}$ signal was measured with an oscilloscope (50 Ω input impedance) and the measured trace was integrated. For the extraction of the pulse energy the $PD_{XUV}$ quantum efficiency as a function of photon energy provided by the manufacturing company (OptoDiode Corp) was used, following the procedures which are comprehensively described elsewhere [23]. When the CCMC-PG was not introduced in the beam path, the energy of the harmonic comb was found to be ~120 μJ per pulse. When the CCMC-PG is implemented and quasi-continuum radiation is generated, a total of ~1.3 μJ per pulse was determined for the whole harmonic spectrum. It should be stressed that these values are referring to the XUV pulse energy at the generation area and their determination is based on the measured $PD_{XUV}$ recordings, taking into account the filter's transmission and the reflectivity of the Si plate.

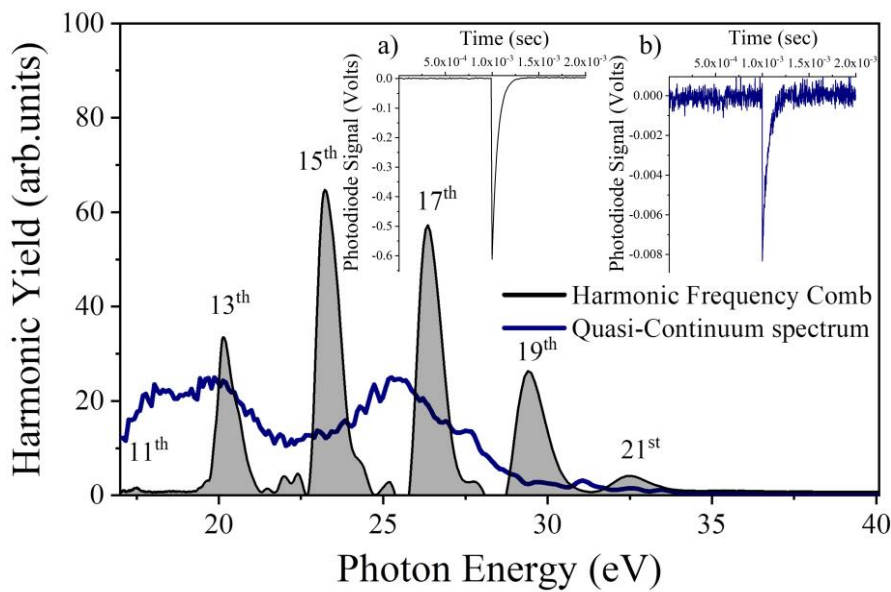

**Fig. 2. Quasi-continuum XUV spectrum**: Black-grey filed solid line is depicting a typical harmonic spectrum generated by a linearly polarized IR pulse consisting of odd HH frequencies. Blue solid line: quasi-continuum XUV spectrum generated by the central part of the elliptically modulated pulse. Each spectrum is an average of 150 shots. The insets (a) and (b) are representing the corresponding pulse energy measurements, with a calibrated $PD_{XUV}$, of the HH (black curve) and quasi-continuum (blue curve) respectively.

Fig. 3 shows the dependence of the XUV spectra and pulse energy on the driving field intensity. The contour plots of Fig.3 (a) and (b) are presenting the harmonic spectra generated in Xenon for the case of the linearly *p*-polarized driving filed, leading to HH, and in the case of ellipticity-modulated IR driving pulse, leading to a quasi-continuum spectrum respectively. As is expected from the electron recollision process, in both cases the harmonic emission cut-off is strongly dependent on the IR intensity. A linear dependence of the highest harmonic order (and photon energies respectively) with the driving intensity is expected, since the cut-off photon energy follows the $E_{cut-off} = I_P + 3.17 U_P$ , where $I_P$ stands for the ionization energy of the non-linear medium and $Up$ is the ponderomotive potential [47], which is the cycle-averaged kinetic energy gained by a free electron oscillating in the radiation field and reads: $U_p(eV) = 9.3 \cdot 10^{-14} \cdot I(\frac{W}{cm^2}) \cdot \lambda^2(\mu m^2)$, where $I$ and $\lambda$ are denoting the intensity and wavelength of the radiation, respectively. Fig. 3 (c) and (d) show in log-log scale the total harmonic yield as a function of the IR intensity in the case of HH frequency comb and quasi-continuum respectively. The highly non-linear nature of the harmonic generation process, is illustrated in the slopes of the straight lines obtained by a linear fit in the data. It is worth noting that in the case of HH the ionization saturation intensity ($I_{sat}$) of Xenon is reached, indicated by the reduction of the slope. The production of the isolated XUV pulses could not be saturated as can be seen in Fig. 3(d). This can be attributed to the dependence of the ionization saturation intensity dependence on the duration of the linearly polarized driving field.

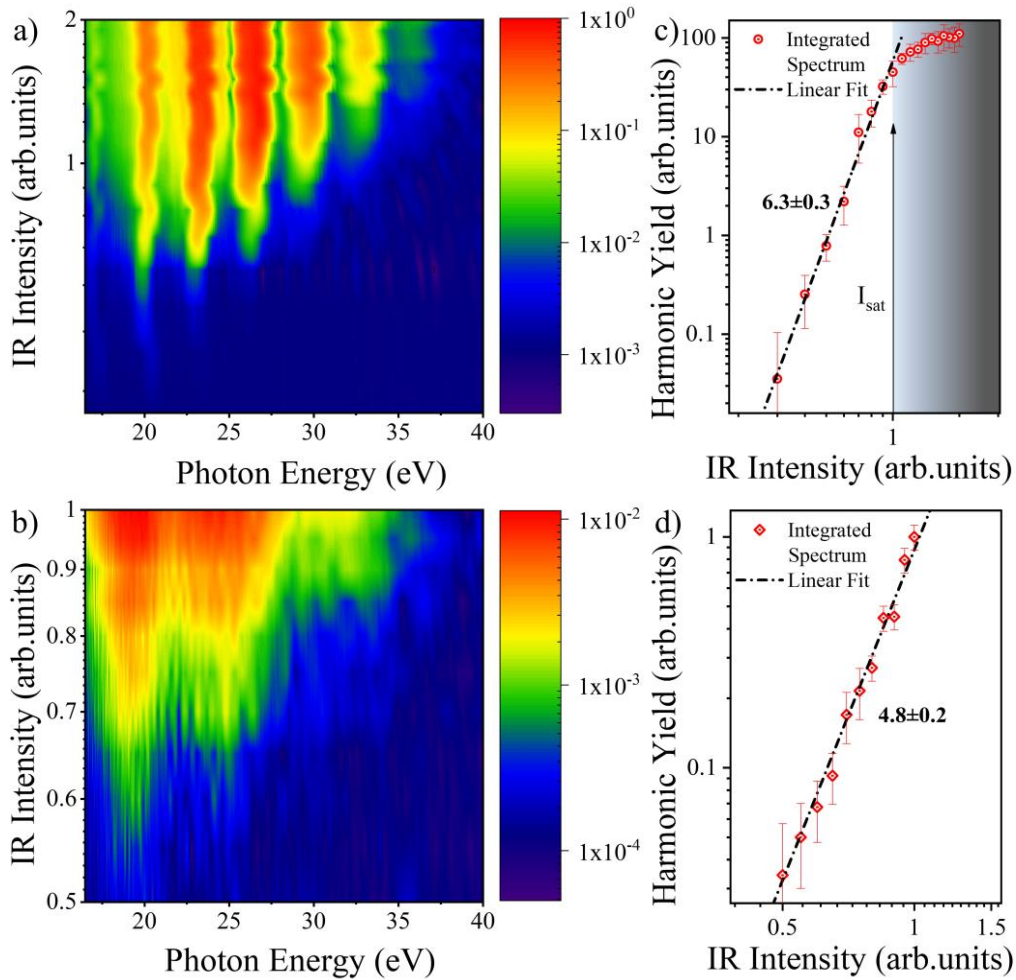

**Fig. 3. XUV generated spectrum dependence on the IR intensity:** (a), (b) Contour plots of the XUV generated spectrum as a function of the IR driving pulse intensity in a semi-log scale in the case of the HH and the quasi-continuum spectrum respectively. (c), (d) Harmonic yield, deduced from the integration of the total spectrum, as a function of the IR intensity in the case of the HH and the quasi-continuum spectrum respectively.

Fig.4 shows the XUV beam profile in the case of the quasi-continuum radiation, generated when the CCMC-PG arrangement is used. For these measurements a $BP_{XUV}$, consisting of a pair MCPs and a phosphor screen followed by a CCD camera mounted on a translation stage, is introduced in the beam path. The $BP_{XUV}$ is placed after the filtering mount and for the recordings a 150 nm thick Al filter is used. As depicted, the single-shot image of the XUV beam profile is exhibiting a Gaussian distribution. Fig. 4 (a) and (b) correspond to the line-outs of the XUV spatial distribution in the horizontal and vertical axis respectively. Fitting a Gaussian function over the measured experimental data a value of ~3 mm full width half maximum (FWHM) is determined.

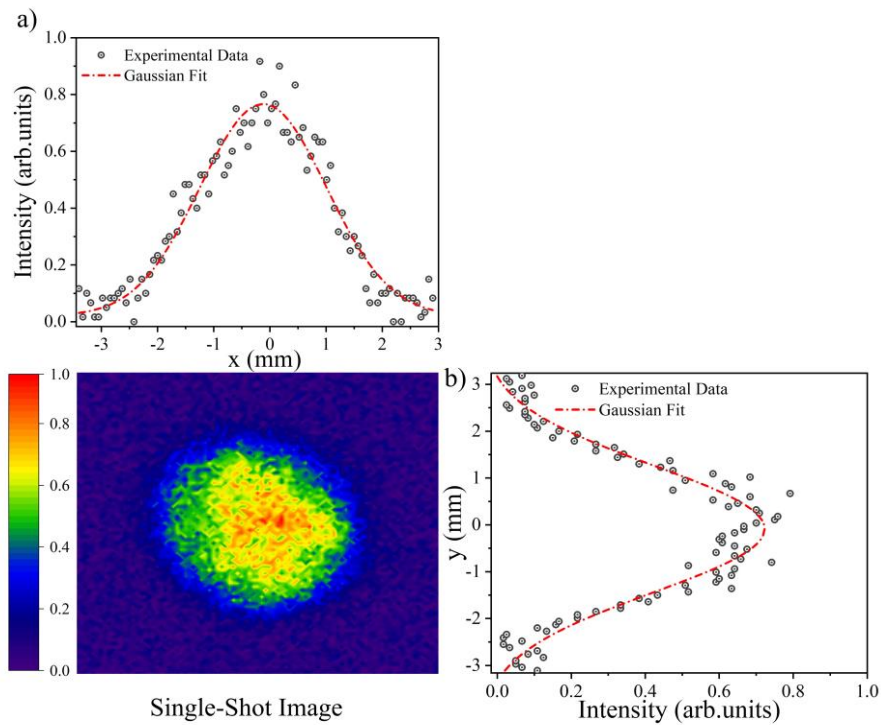

**Fig. 4. XUV beam profile of the quasi-continuum XUV:** A single shot image of the spatial distribution of the quasi-continuum XUV radiation, recorded by means of an XUV beam profiler ($BP_{XUV}$). (a) and (b) are depicting the line-outs of the distributions along the x and y-axis respectively.

## 3. Theoretical calculations

In the process of generating isolated *as* pulses by a few-cycle laser system, the CEP of the driving field is a decisive parameter **[48]**. The same holds in the polarization gating approaches **[44,49]**. Thus, as the driving fields in the present work are not CEP stable, only a fraction of the delivered shots of the harmonic radiation are isolated *as* pulses. As shown in Fig.5 (a), the generated XUV spectra alternate from a pure continuum structure to a modulated continuum featuring discrete features as the CEP of the driving pulse is fluctuating. These variations of the generated spectrum lead to different temporal profiles of the emitted XUV pulses. The insets of Fig.5(a) (i) and (b) (ii) show the pulse durations derived (for the unrealistic case of Fourier Transform Limited (FTL) XUV pulses) from the Fourier transform of the continuum (blue curve) and modulated spectra (red dashed-dot curve). A single isolated *as* pulse with minimum duration of~ 350 *as* and a narrow APT of ~ 2 *fs* duration are obtained respectively.

As presented in the three characteristic different single shot recordings of Fig.5 (a), due to the dependence of harmonic peak position shift on the CEP, the generated spectra undergo variations regarding their structure and frequency shift. A frequency shift *Δω* of the harmonic peak's structure is illustrated in the case where spectra with discrete features are generated, while a spectrum exhibiting an almost pure continuum structure is also present. This behavior is evidencing the influence of the CEP drift in the current experiment. The presence of a narrow gate width with linear polarization (where the XUV emission takes place) at the center of the ellipticity-modulated IR pulse and the dependence of the harmonic peak position on the CEP can be exploited in determining the CEP value [44]. Assuming a cosine-type driving field with CEP equal to π/2 a continuum emission, corresponding to an IAP, is occurring. On the contrary, for CEP values different than π/2, harmonic fields emitted from time intervals greater than half a laser period lead to formation of APT i.e. to discrete harmonic spectrum. In the extreme cases of CEP values of 0 and π the harmonic peaks are exhibiting their maximum frequency shift while for the in-between values the dependence of the harmonic shift on the CEP is linear [44,49]. Hence, the generation of a continuum spectrum (blue solid curve in Fig.5 (a)) corresponds to a CEP value of ≈π/2, while modulated spectra (black doted and red dashed dot curves of Fig.5 (a)), corresponds to ≈0 and ≈π CEP.

The above have been confirmed by theoretical simulations based on the solution of the well-known semi-classical three step model [50]. The ellipticity-modulated field is reconstructed taking into account the propagation of the pulse through the CCMC-PG system. The gate width, i.e., time interval within which the harmonic generation is essentially occurring, was is $\tau_g$=1.5 *fs*. The gate width is quantified from the variation of the ellipticity *ε* from zero to the value of $\varepsilon_{th}$, which corresponds to the level that the harmonic efficiency decreases to half. The intensity within the gate is $I_g$=0.7×10$^{14}$ W/cm$^2$. With these parameters of the driving laser field, the XUV spectrum generated by Xenon atoms has been calculated. After calculating the acceleration of the single-atom dipole moment $x(t)$ and its Fourier transformation $x(\omega)$ the harmonic emission rate $W(\omega) = \omega^3|x(\omega)|^2$ is obtained. In the present work both quantum trajectories, namely short and long, of the electron accelerated in the continuum have been taking into account, owing equal contribution to the total emitted harmonic spectrum. This assumption was strongly motivated by the experimental conditions under which the quasi-continuum radiation is produced. In particular, as mentioned in the experimental part, a Xenon gas jet was placed at the focus of the driving IR beam, condition favoring both electron trajectories almost with equal weight. Thus, in the calculated spectra both electron trajectories, as well as their interference in the total emission have been considered, with the phase of the radiation to be the average of the independently accumulated electron trajectories. In Fig.5 (b) the calculated spectrum (red dashed-dot line) is presented along with the experimentally obtained one (blue solid line) for comparison purposes. The experimental spectrum is acquired from the average over 150 shots. The inset (i) of Fig.5b is showing the Fourier transform of the calculated XUV spectrum, while (ii) shows the average temporal profile, calculated for several different values of the CEP ($0 \leq CEP \leq \pi$).

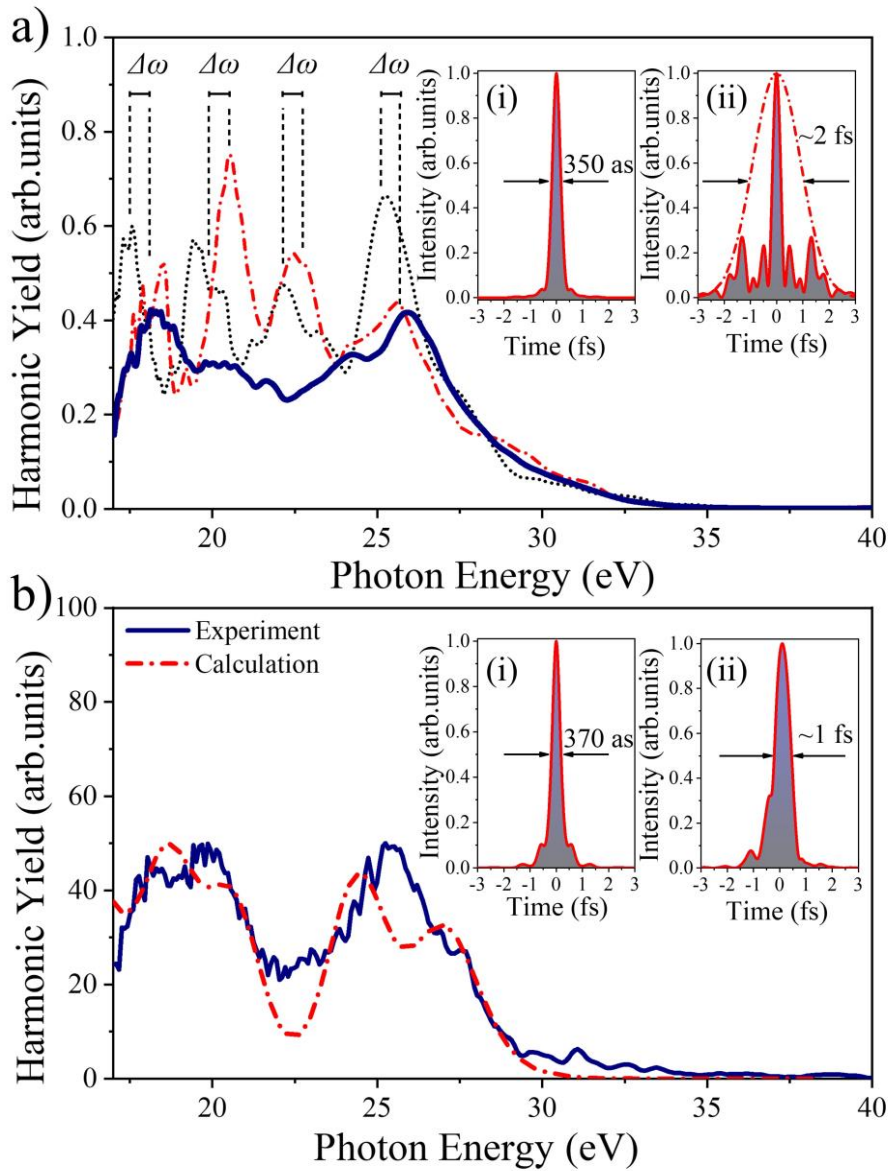

**Fig. 5. Measurement of the CEP variation and calculated XUV spectrum**: (a) single-shot XUV spectra for different CEP values evidenced by the harmonic frequency shift $\Delta\omega$ (black dot and red dashed-dot curves), along with an XUV spectrum exhibiting a continuum structure (blue curve). The insets a) (i) and a) (ii) show the pulse durations

derived from the Fourier Transform of the modulated (red dashed-dot curve) and continuum spectra respectively (blue curve). (b) the blue line shows a typical measured XUV spectrum, accumulating 150 shots of average. The red dash-dot line is presenting the calculated XUV spectrum obtained by solving the single atom, semi-classical three step model for an ellipticity-modulated field, taking into account the present experimental conditions. The inset b) (i) shows the Fourier Transform of the calculated XUV spectrum, while b) (ii) shows the average temporal profile of the XUV pulse calculated over several different values of the CEP ($0 \leq CEP \leq \pi$) of the driving field.

## 4. Two-XUV-photon double ionization of Argon

The XUV pulses generated in Xenon gas using the CCMC-PG system, were focused by a spherical Au coated mirror into an Argon gas jet. The ions produced by the interaction of the focused XUV beam with Argon atoms have been recorded by a MB-TOF ion mass spectrometer. Taking into account the measured XUV energy, the measured ~1.5 % overall XUV transmission factor of the beam line and the ~12 % reflectivity of the Au spherical mirror [19], the energy of the harmonic radiation in the interaction region has been estimated to be ~20 nJ. Considering a focal spot diameter of 2 μm [19] and an average XUV pulse duration of ~1 *fs* it is estimated that the intensity of the XUV pulses in the interaction area can be up to $10^{14}$ W/cm$^2$. Such XUV intensities are sufficient to induce non-linear processes in Argon atoms and lead to the production of $Ar^{2+}$ via two-photon double ionization process. Fig. 6 (a) shows the excitation scheme of this process. There are two possible channels producing $Ar^{2+}$ ions. the direct, through simultaneous absorption of two XUV photons, or sequential where a single photon leads to the production of $Ar^+$ which subsequently absorbs a second photon towards the production of $Ar^{2+}$.

The production of $Ar^+$ and $Ar^{2+}$ is shown in the TOF mass spectrum of Fig. 6(b). The other peaks can be attributed to rest gas and contamination species of the chamber and of the gas valve. Those include, $H^+$, $N^+$, $O^+$, $H_2O^+$, $N_2^+$ and $O_2^+$ and they are produced by single photon ionization processes since the photoexcitation energy is exceeding their ionization energies. The inset of Fig. 6(b) shows an expanded area of the portion of TOF spectrum, in which $Ar^{2+}$ is clearly visible.

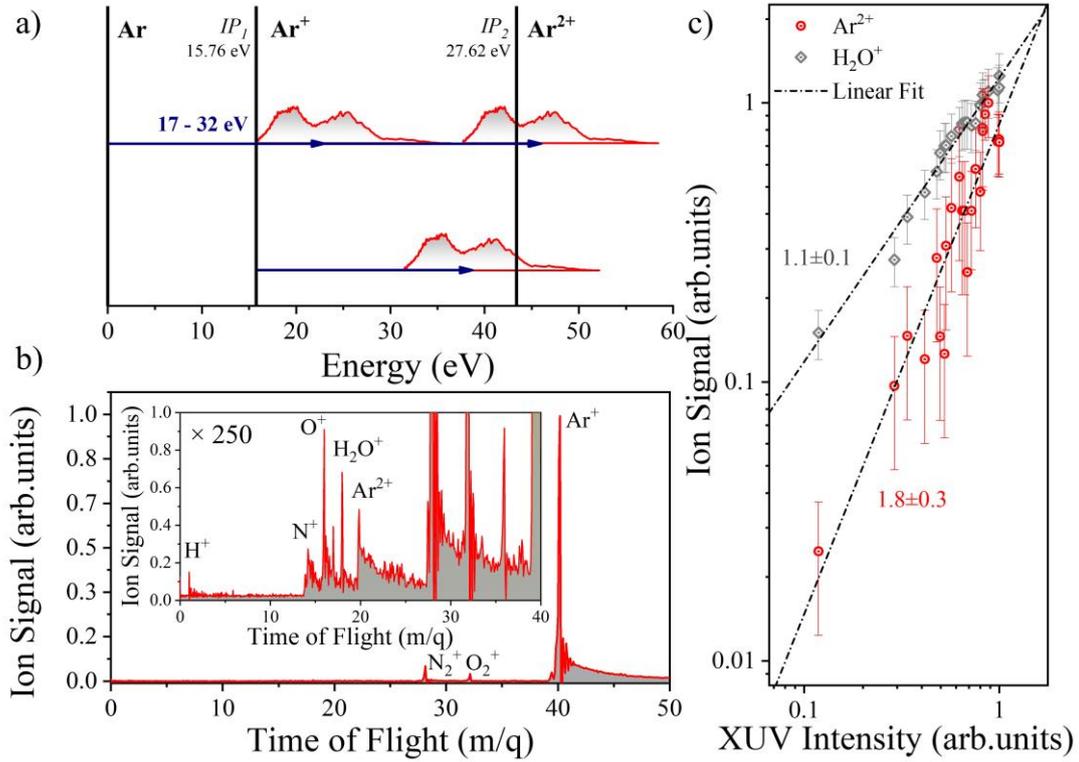

**Fig. 6. Two-XUV-photon double ionization of Argon induced by the intense quasi-continuum XUV radiation:** (a) Two-XUV-photon double ionization scheme of Argon. $Ar^{2+}$ is formed through either a direct or a sequential ionization process. (b) TOF mass spectrum produced by the interaction of the focused quasi-continuum XUV radiation with Argon. (c) Dependence of $Ar^{2+}$ and $H_2O^+$ yields on the $I_{XUV}$. The slope of 1.8±0.3 in the case of $Ar^{2+}$ is evidencing the second order non-linearity of the ionization process, while $H_2O^+$ ion yield has a linear dependence on the XUV intensity, with a slope of 1.1±0.1, as expected for a single-photon ionization process. The black dashed lines show a linear fit on the $Ar^{2+}$ and $H_2O^+$ data.

In order to confirm the non-linear nature of the process the dependence of the $Ar^{2+}$ yield on the XUV intensity has been measured (Fig. 6c). The measured quadratic behavior is indicative of a second order non-linear process, as expected according to lowest-order perturbation theory. In particular, one expects the ion yield produced via a non-linear process to scale proportionally with $I^n$, with $I$ being the intensity of radiation and $n$ the minimum number of the photons required by the ionization process [28]. The XUV pulse energy was varied by changing the delay between the opening of the HHG gas-jet nozzle and the arrival time of the driving IR pulse, thus changing the effective atomic density within the time interval of the interaction between the IR pulse and the gas medium. This controllable variation of the total XUV energy is illustrated by the linear variation of the yields of the single photon ionization products, such as the

$H_2O^+$ which is presented along with $Ar^{2+}$ in Fig. 6(c). The measured yield of single photon ionization products is proportional to the XUV intensity in the interaction region.

A quantitative investigation on which of these channels is dominant requires the solution of the TDSE of Ar atoms in the presence of a broadband XUV field, which is out of the scope of the present work. However, at the present experimental conditions the pulse duration reaches to the *fs-as* barrier **[51]**, assuming a ~ 1 *fs* XUV pulse. Moreover, in the particular case of the current experiment, the excitation photon energies are fulfilling the condition $\frac{IP_2}{2} \leq \hbar\omega \leq IP_2 - IP_1$, where *IP₁* and *IP₂* stands for the first and second ionization thresholds of the atomic ground state respectively. Under such conditions, the direct ionization process can become the dominant one, as demonstrated in previous studies with a similar excitation scheme **[27,30]**. Taking into account the ionization energies of Ar and $Ar^+$, as well as the photon energy of the quasi-continuum XUV pulses in the current experimental studies this condition is satisfied by the largest part of the ionizing spectrum as depicted in Fig. 6 (a). Thus, the formation of $Ar^{2+}$ via a sequential pathway, would require a three-photon sequential double ionization process (ThPSDI) with the photoionization rate reading: $\frac{dP_{ThPSDI}}{dt} = \sigma_{Ar}^{(1)} \sigma_{Ar^+}^{(2)} \left(\frac{I_{XUV}}{\hbar\omega}\right)^3 \tau$. Therefore, the direct channel leading to double ionization would be favored by the lower order of nonlinearity required compared to that of the sequential one. In fact, the measured slope of 1.8±0.3 is clearly evidencing a second order process rather than a sequential three-photon one. This could be resulted either by a simultaneous absorption of two XUV photons from the ground state of Argon or with a sequential excitation process in which the single photon ionization is saturated. In the latter case at intensities approaching the saturation conditions of the atomic ground state, both processes are expected to contribute substantially with the different rates of ionization resulting in a modification of the ion yield with respect to intensity **[52]**.

## 5. Conclusions and perspectives

In conclusion, in the present work micro-Joule level of coherent quasi-continuum XUV radiation generated by a multi-cycle Laser field exploiting a polarization gating technique is presented. The production of the broadband energetic emission is achieved by implementing a compact-collinear, many-cycle-polarization gating technique. Theoretical calculations support the observed results and allowed to estimate the XUV pulse duration which, depending on the CEP value, found to be in the range from 370 *as* to 1.5 *fs*. The total generated pulse energy of ~ 1.3 μJ in conduction with the short pulse duration upon focus can reach intensities on the order of $10^{14}$ W/cm$^2$, sufficient to induce non-linear processes in the XUV spectral region. Two-XUV double ionization of Argon induced by the quasi-continuum XUV radiation is observed and studied, conducting XUV power dependence measurements, revealing the non-linear nature of the ionization process. The results of this work demonstrate the capability of the recently developed GW attosecond source at FO.R.T.H. of producing intense coherent quasi-continuum XUV radiation supporting isolated *as* pules. It also addresses the capability of the beamline of conducting XUV-pump–XUV-probe experiments using isolated *as* pulses. Such processes can be exploited in attosecond pulse metrology as well as in the studies of the ultrafast evolving atomic coherences.

## Acknowledgments

We acknowledge support of this work by the Hellenic Foundation for Research and Innovation (HFRI) and the General Secretariat for Research and Technology (GSRT) under grant agreements [GAICPEU (Grant No 645)] and NEA-APS HFRI-FM17-3173, the H2020 project IMPULSE (GA 871161). The ELI-ALPS Project (GINOP-2.3.6-15-2015-00001) is supported by the European Union and it is co-financed by the European Regional Development Fund.